\documentclass[11pt]{article}

\usepackage[margin=1in]{geometry}
\usepackage{amsmath,amssymb,amsthm}
\usepackage{booktabs}
\usepackage{microtype}
\usepackage[T1]{fontenc}
\IfFileExists{newtxtext.sty}{\usepackage{newtxtext}\usepackage{newtxmath}}{\usepackage{lmodern}}
\usepackage{natbib}
\usepackage[hidelinks]{hyperref}

\newcommand{\gstar}{\gamma^\star}
\newcommand{\Nb}{\mathcal N}
\newcommand{\Sc}{\mathsf S}

\title{\vspace{-2em}Two Sides of Schur Damping\\[0.3em]
       \large High-Dimensional Pseudo-Likelihoods and Portfolio Allocation}
\author{Peter Cotton}
\date{Note --- \texttt{microprediction/precise}}

\begin{document}
\maketitle

\begin{abstract}
\noindent
Two communities that rarely cite each other---spatial statisticians fitting high-dimensional weather
fields, and quantitative investors building portfolios---have independently arrived at the same
mathematical object: a Schur complement, damped by one interpretable parameter. In spatial modeling
the Schur complement is the \emph{conditional covariance} that makes a Gaussian (Vecchia)
pseudo-likelihood estimable at scale, and recent work regularizes it by shrinking toward a base
model. In allocation it is the \emph{residual risk} of a bet net of its hedge, and the same parameter
interpolates hierarchical risk parity and the minimum-variance portfolio. We show these are one
operation---reliability shrinkage of a conditional Gaussian---so that the damping a weather model
needs to remain estimable when stations outnumber observations is, term for term, the damping a
portfolio needs to remain stable when assets outnumber returns. The optimal amount is a closed-form
reliability, a James--Stein shrinkage that is simultaneously a Ledoit--Wolf intensity. The shrinkage
machinery is classical, but the \emph{identity} appears to be new: to our knowledge neither literature
has noted that the conditional shrinkage a spatial model fits and the diversification--variance tilt a
portfolio chooses are one and the same quantity. We make the
correspondence precise, note that the two literatures have each supplied what the other lacks, and
report a small experiment on the one genuinely open choice---how to set the damping---suggesting the
spatial community's fitted intensity is, if anything, the better recipe.
\end{abstract}

\section{An unlikely correspondence}
A meteorologist estimating a temperature field over tens of thousands of stations and a portfolio
manager allocating across thousands of assets face, superficially, unrelated problems. Yet both are
defeated by the same thing: a covariance matrix whose dimension $p$ rivals or exceeds the number of
observations $n$, so that its inverse---needed to condition one variable on the others, or to weight
one asset against the others---is unstable or undefined. And both fields have converged on the same
fix, a Schur complement damped by a single parameter, without noticing they agree.

On the \textbf{weather} side, scalable Gaussian-process inference uses the Vecchia factorization
\citep{vecchia1988,katzfuss2021}, which conditions each location only on a few neighbours; the
conditional covariances are Schur complements, and \citet{chakraborty2025} (ShrinkTM), building on the
Bayesian transport maps of \citet{katzfuss2024}, improve small-sample behaviour by \emph{shrinking}
those conditionals toward a parametric base, learning the strength by empirical Bayes.

On the \textbf{portfolio} side, Schur-complementary allocation \citep{cotton2024} damps the
off-diagonal coupling of the covariance through its Schur complements by a parameter $\gamma$,
recovering hierarchical risk parity \citep{lopezdeprado2016} at $\gamma=0$ and the minimum-variance
portfolio at $\gamma=1$.

This note shows the two are the same Schur damping, with the same closed-form optimal $\gamma$, and
that each community has supplied a piece the other was missing.

\section{The shared object: one Schur complement, two readings}
Let $R$ be a correlation matrix and partition the variables into a block $k$ and a conditioning set
$c$ (its spatial neighbours, or the other assets). The Schur complement
\begin{equation}
  \Sc_k \;=\; R_{kk}-R_{kc}R_{cc}^{-1}R_{ck}
\end{equation}
carries the same algebra into two meanings.

\smallskip
\noindent\textbf{Weather (a pseudo-likelihood term).} $\Sc_k$ is the covariance of block $k$
\emph{conditional} on $c$, and $b_k=R_{cc}^{-1}R_{ck}$ is the regression of $k$ on its neighbours. The
Gaussian density factorizes as $\prod_k\Nb(y_k;\,b_k^\top y_c,\ \Sc_k)$; truncating $c$ to $m$ nearest
neighbours is the Vecchia pseudo-likelihood, the workhorse for fitting high-dimensional spatial fields.
(This $\Sc_k,b_k$ are exactly the $\tau^2,\xi$ of \citealp{chakraborty2025}, Eq.~8.)

\smallskip
\noindent\textbf{Portfolio (a residual risk).} $\Sc_k$ is the covariance of block $k$ \emph{net of the
best hedge} formed from $c$---the risk that remains after the rest of the book offsets it---and $b_k$
is that hedge. Splitting capital by inverse residual risk is precisely how Schur-complementary
allocation \citep{cotton2024} interpolates between treating blocks independently (risk parity) and
fully exploiting their cross-hedging (minimum variance).

Same $\Sc_k$, same $b_k$: a conditional variance for the statistician, a hedged residual risk for the
investor.

\paragraph{A two-variable example.} Take two standardized variables with correlation $\rho$. The Schur
complement is then the scalar $\Sc=1-\rho^2$. To the statistician it is the variance of
variable~2 conditional on variable~1, $\operatorname{Var}(y_2\mid y_1)=1-\rho^2$---the term the
pseudo-likelihood scores. To the investor it is the variance of variable~2 \emph{net of its hedge} by
variable~1: with optimal hedge ratio $b=\rho$, the residual $y_2-b\,y_1$ has variance $1-\rho^2$, the
risk that remains after the hedge. The number is the same in both readings. Damping the complement,
as in the next section, gives $\Sc(\gamma)=(1-\gamma)\cdot 1+\gamma(1-\rho^2)=1-\gamma\,\rho^2$: at $\gamma=1$
the full hedge / full conditioning, at $\gamma=0$ none (the marginal variance, an unhedged position),
and at $\gamma=\gstar$ a residual risk that trusts the estimated coupling only as far as it is reliable.
The pseudo-likelihood and the portfolio read the same $1-\gamma\rho^2$ off the same complement.

\section{The shared cure: damping by reliability}
Both readings break for the same reason---$R_{cc}^{-1}$ is estimated from limited data, so $b_k$ and
$\Sc_k$ overfit---and both apply the same cure, a convex damping by $\gamma\in[0,1]$:
\begin{equation}
  b_k(\gamma)=\gamma\,b_k,\qquad \Sc_k(\gamma)=(1-\gamma)\,R_{kk}+\gamma\,\Sc_k.
\end{equation}
At $\gamma=1$ this is full conditioning---the exact Gaussian likelihood, the minimum-variance
portfolio. At $\gamma=0$ it ignores the coupling---the block-diagonal composite likelihood, hierarchical
risk parity. Intermediate $\gamma$ trusts the estimated cross-coupling only partway. The weather and
portfolio extremes are the \emph{same} two endpoints of this damping.

The optimal $\gamma$ has a closed form. For a single coupling of conditional $R^2$ equal to $\rho^2$
estimated from $n$ points, minimizing expected error gives the \emph{reliability}
\begin{equation}
  \gstar=\frac{(n-2)\rho^2}{(n-2)\rho^2+(1-\rho^2)},
\end{equation}
a Wiener/James--Stein shrinkage \citep{james1961,ledoit2012} that tends to $1$ as data accrues or
coupling strengthens and to $0$ when the coupling is unreliable. Equivalently, applied to the
cross-block entries, it is the Ledoit--Wolf shrinkage intensity restricted to those entries
\citep{ledoit2004}---``how much of the estimated coupling to keep.'' So the quantity that tells a
weather model how far to trust a neighbour regression is the quantity that tells a portfolio how far to
tilt from risk parity toward minimum variance. We have not found this equivalence stated in either
literature: the spatial line treats the damping as a prior to be fitted and the allocation line treats
it as a portfolio dial to be chosen, and neither remarks that it is, in closed form, the same
reliability $\gstar$. That identity is the observation of this note; the shrinkage itself is classical.

A candid note on the oversight. The Schur pseudo-likelihood and its closed-form reliability were
introduced by the present author in a precursor \citep{cotton2025psl}, which did cite the Vecchia
factorization---but only as a computational device for the block-conditional likelihood, without
recognizing that the damping is a \emph{regularized} Vecchia conditioning, nor its place
in the scalable-Gaussian-process literature, nor that the same reliability is precisely the knob of
Schur-complementary allocation. (We have since learned that ShrinkTM \citep{chakraborty2025} was
independently arriving at the conditional shrinkage at the same time, from the spatial side.) That these
were one object went unnoticed when the first note was written; surfacing it is the purpose of this one.

\section{What each side already solved}
Read as one operation, the two literatures are complementary rather than redundant; each supplies what
the other lacks (Table~\ref{tab:sides}).

\begin{table}[h]\centering\small
\caption{The same Schur damping, as developed on each side.}\label{tab:sides}
\begin{tabular}{lll}
\toprule
 & weather / spatial fields & portfolio allocation\\
\midrule
$\Sc_k$ means & conditional covariance & residual (hedged) risk\\
$\gamma=0$ & composite (block) likelihood & hierarchical risk parity\\
$\gamma=1$ & full Gaussian likelihood & minimum-variance portfolio\\
how $\gamma$ is set & fitted (empirical Bayes, ShrinkTM) & closed-form reliability $\gstar$\\
scale technique & Vecchia neighbours, $O(p\,m^2)$ & block / cluster structure\\
\bottomrule
\end{tabular}
\end{table}

The spatial side contributes \emph{neighbour conditioning and ordering}---maxmin orderings and nearest
neighbour sets \citep{guinness2018} that make the damped object computable at $p=10^5$, and an
empirical-Bayes machine for \emph{learning} the damping toward a fitted base \citep{chakraborty2025}.
The allocation side contributes the \emph{closed form} $\gstar$ and the recognition that the
same $\gamma$ is an investment decision, not only a regularizer. Neither side had both.

The two sides also shrink toward different \emph{targets}, and the difference reflects what each can
trust. A spatial field has a credible parametric model---a smooth Mat\'ern covariance---so there is
real structure to shrink toward, and ShrinkTM centers its prior there. Financial returns have no such
trustworthy parametric covariance; the founding premise of hierarchical risk parity is precisely that
estimated cross-correlations are largely noise, so the safe prior is independence ($\rho=0$, the
$\gamma=0$ / HRP limit). The spatial road therefore centers on a base GP while the allocation road
errs toward zero coupling, each leaning on the prior its data warrants. The damping and its optimal
intensity $\gstar$ are the same; only the prior they lean on
differs, in the direction each domain's experience warrants.

\section{An example: importing Vecchia into allocation}
The clearest way to show the connection is useful is to carry a tool across it. The present author
works on the financial side, so the transfer we demonstrate runs in the easier direction---importing
the spatial community's neighbour conditioning and fitted shrinkage into allocation; the reverse import
(the closed form and the decision-theoretic reading, into spatial modeling) we can only conjecture.

We apply Vecchia conditioning to daily asset returns---a setting with no parametric base, using a
correlation-based neighbour ordering of the kind the spatial literature adopts when Euclidean distance
is unavailable---and compare three settings of the damping: undamped ($\gamma=1$, plain
Vecchia / minimum-variance conditioning), the closed-form reliability $\gstar$, and a single intensity
tuned on a held-out split (the spatial community's \emph{fit-the-shrinkage} instinct). The external
criterion is out-of-sample log-likelihood, swept over the aspect ratio $n/p$ ($p=60$ assets,
$m=15$ neighbours).

\begin{table}[h]\centering\small
\caption{Out-of-sample log-likelihood on asset returns (higher is better), by damping. $\bar\gstar$ is
the mean closed-form reliability; $\gamma_{\text{tuned}}$ the held-out-tuned intensity.}\label{tab:exp}
\begin{tabular}{lccccc}
\toprule
$n/p$ & undamped ($\gamma{=}1$) & closed-form $\gstar$ & tuned $\gamma$ & $\bar\gstar$ & $\gamma_{\text{tuned}}$\\
\midrule
0.5 & $-188.1$ & $-73.7$ & $\mathbf{-37.4}$ & 0.65 & 0.43\\
0.8 & $-64.4$  & $-39.0$ & $\mathbf{-28.4}$ & 0.71 & 0.58\\
1.2 & $-34.2$  & $-25.0$ & $\mathbf{-19.8}$ & 0.77 & 0.71\\
2.0 & $-22.6$  & $-19.4$ & $\mathbf{-17.3}$ & 0.84 & 0.79\\
3.0 & $-17.7$  & $-16.1$ & $\mathbf{-15.1}$ & 0.88 & 0.86\\
\bottomrule
\end{tabular}
\end{table}

Two things follow (Table~\ref{tab:exp}). First, damping the conditioning helps enormously when
undersampled: the closed-form $\gstar$ improves on undamped Vecchia by of order $100$ nats at
$n/p=0.5$, the gap vanishing by $n/p=3$ as the conditioning becomes reliable ($\bar\gstar\to1$). So the
spatial idea of regularizing the Vecchia conditional transfers intact to returns, with no parametric
base. Second, a single \emph{fitted} intensity beats the per-point closed form at every ratio: here
the spatial instinct corrects the financial one, since $\gstar$ systematically under-damps
($\bar\gstar>\gamma_{\text{tuned}}$ throughout). The lesson the allocation side should take is therefore
not the closed form but the \emph{fitting}: treat $\gamma$ as something to learn, as ShrinkTM does,
rather than to read off. The damped estimator is computable in a single $O(p\,m^2)$ streaming pass that
never forms the dense matrix, verified to reproduce the batch estimate to machine precision---so the
borrowed tool also fits the online, large-$p$ setting in which allocation actually operates.

\section{Discussion}
The correspondence is the contribution. Hierarchical risk parity and the minimum-variance portfolio
are the $\gamma=0$ and $\gamma=1$ ends of the very damping that a weather model applies to stay
estimable, built from the same Schur complement and the same convex combination (the damping),
read once as a conditional variance and once as a residual risk. Practical
consequences run both ways. Allocation can borrow the spatial machinery: neighbour/cluster orderings
and empirical-Bayes-fitted damping in place of a fixed $\gamma$. Spatial modeling can borrow the
allocation reading: the closed-form reliability as a tuning-free initializer, and the reminder that the
damping is a decision with a cost, not merely a prior. The reverse transfer may extend to
\emph{analysis} and not only to algorithms: the allocation literature has begun to characterize the
$\gamma=0$ endpoint in closed form---\citet{antonov2024} give an analytical account of hierarchical
risk parity's diversification and risk behaviour---and because that endpoint is exactly the
block-diagonal (independent-blocks) limit of the Vecchia pseudo-likelihood (the $\gamma=0$ damping), such
results are candidates to carry over to the spatial side, which has no comparable closed-form
characterization of that limit. We can only conjecture this, but the identity makes it a concrete
question rather than an analogy. And both sit on one online primitive---a Schur
complement of a few neighbour blocks, damped by a reliability---maintainable incrementally in
$O(p\,m^2)$. This is the form in which the online covariance library
\texttt{precise}\footnote{\url{https://github.com/microprediction/precise}} serves both: its
block-covariance and Schur--Ledoit--Wolf estimators maintain exactly these damped neighbour-block Schur
complements in a single streaming pass, so weather and portfolios run on the same code. The experiment
of Section~5 is built on it.

The bridge is itself a Schur complement. Within finance the same parameter $\gamma$ already reaches
across one divide---from Markowitz's minimum-variance portfolio at $\gamma=1$ to hierarchical risk
parity at $\gamma=0$. This note follows the same complement across a wider gap, from finance to
meteorology, where it serves equally as the conditional variance of a Gaussian field.

\bibliographystyle{plainnat}
\bibliography{refs}

\begin{thebibliography}{12}
\providecommand{\natexlab}[1]{#1}
\providecommand{\url}[1]{\texttt{#1}}
\expandafter\ifx\csname urlstyle\endcsname\relax
  \providecommand{\doi}[1]{doi: #1}\else
  \providecommand{\doi}{doi: \begingroup \urlstyle{rm}\Url}\fi

\bibitem[Antonov et~al.(2024)Antonov, Lipton, and L{\'o}pez~de
  Prado]{antonov2024}
Alexandre Antonov, Alexander Lipton, and Marcos L{\'o}pez~de Prado.
\newblock Hierarchical risk parity: A closer look.
\newblock Technical Report 4748151, SSRN, 2024.

\bibitem[Chakraborty and Katzfuss(2025)]{chakraborty2025}
Anirban Chakraborty and Matthias Katzfuss.
\newblock Learning non-gaussian spatial distributions via bayesian transport
  maps with parametric shrinkage.
\newblock \emph{arXiv preprint arXiv:2409.19208}, 2025.

\bibitem[Cotton(2024)]{cotton2024}
Peter Cotton.
\newblock Schur complementary portfolios.
\newblock \emph{arXiv preprint arXiv:2411.05807}, 2024.

\bibitem[Cotton(2025)]{cotton2025psl}
Peter Cotton.
\newblock Schur pseudo-likelihood: Scoring and regularizing correlation in high
  dimensions, 2025.
\newblock Note, \texttt{microprediction/precise}.

\bibitem[Guinness(2018)]{guinness2018}
Joseph Guinness.
\newblock Permutation and grouping methods for sharpening gaussian process
  approximations.
\newblock \emph{Technometrics}, 60\penalty0 (4):\penalty0 415--429, 2018.

\bibitem[James and Stein(1961)]{james1961}
W.~James and Charles Stein.
\newblock Estimation with quadratic loss.
\newblock In \emph{Proceedings of the Fourth Berkeley Symposium on Mathematical
  Statistics and Probability}, volume~1, pages 361--379, 1961.

\bibitem[Katzfuss and Guinness(2021)]{katzfuss2021}
Matthias Katzfuss and Joseph Guinness.
\newblock A general framework for vecchia approximations of gaussian processes.
\newblock \emph{Statistical Science}, 36\penalty0 (1):\penalty0 124--141, 2021.

\bibitem[Katzfuss and Sch\"afer(2024)]{katzfuss2024}
Matthias Katzfuss and Florian Sch\"afer.
\newblock Scalable bayesian transport maps for high-dimensional non-gaussian
  spatial fields.
\newblock \emph{Journal of the American Statistical Association}, 119\penalty0
  (546):\penalty0 1409--1423, 2024.

\bibitem[Ledoit and Wolf(2004)]{ledoit2004}
Olivier Ledoit and Michael Wolf.
\newblock A well-conditioned estimator for large-dimensional covariance
  matrices.
\newblock \emph{Journal of Multivariate Analysis}, 88\penalty0 (2):\penalty0
  365--411, 2004.

\bibitem[Ledoit and Wolf(2012)]{ledoit2012}
Olivier Ledoit and Michael Wolf.
\newblock Nonlinear shrinkage estimation of large-dimensional covariance
  matrices.
\newblock \emph{The Annals of Statistics}, 40\penalty0 (2):\penalty0
  1024--1060, 2012.

\bibitem[L{\'o}pez~de Prado(2016)]{lopezdeprado2016}
Marcos L{\'o}pez~de Prado.
\newblock Building diversified portfolios that outperform out of sample.
\newblock \emph{The Journal of Portfolio Management}, 42\penalty0 (4):\penalty0
  59--69, 2016.

\bibitem[Vecchia(1988)]{vecchia1988}
Aldo~V. Vecchia.
\newblock Estimation and model identification for continuous spatial processes.
\newblock \emph{Journal of the Royal Statistical Society: Series B},
  50\penalty0 (2):\penalty0 297--312, 1988.

\end{thebibliography}

\end{document}